\documentclass[letterpaper, 10 pt, journal,twocolumn]{IEEEtran}
\IEEEoverridecommandlockouts   
\usepackage{cite}
\usepackage{mathtools,cuted}
\usepackage{amsmath,amssymb,amsfonts}
\usepackage{graphicx}
\usepackage{textcomp}
\usepackage{xcolor}
\usepackage{hyperref}
\usepackage{booktabs}
\usepackage{multirow}
\usepackage{tabularx}
\usepackage{courier}
\usepackage{float}
\usepackage{makecell}
\setcellgapes{5pt}
\usepackage{subcaption}
\UseRawInputEncoding
\usepackage{comment}

\usepackage{amssymb, amsmath, amsfonts}
\usepackage{ graphicx, epstopdf}
\usepackage{algorithm,  multirow,titlecaps,algorithmicx} 
\usepackage[T1]{fontenc}
\usepackage{marvosym}

\usepackage{tikzsymbols}
\DeclareMathAlphabet{\mathbbold}{U}{bbold}{m}{n}

\usepackage[english]{babel}
\usepackage[utf8]{inputenc}
\usepackage[noend]{algpseudocode} 
\usepackage{amsmath,amssymb,amsfonts}
\usepackage{graphicx}
\usepackage{xcolor}
\usepackage{booktabs}
\usepackage{multirow}
\usepackage{courier}
\usepackage{float}
\usepackage{makecell}
\setcellgapes{5pt}
\newcommand{\ra}[1]{\renewcommand{\arraystretch}{#1}}

\newcommand{\cmr}[1]{{\color{red}{#1}}}

\newcommand{\stiny}[1]{{\scalebox{.5}{#1}}}

\newcommand{\argmax}{{\mathrm{argmax}}}

\newcommand*{\minOp}{\operatornamewithlimits{min}\limits}

\newcommand{\tr}{{\mathsf{T}}}

\iftrue
\newcommand{\eye}{\mathbb{I}}
\else
\newcommand{\eye}{\mathbf{I}}
\fi

\newcommand{\vc}[1]{{ \mathrm{#1} }}
\newcommand{\mx}[1]{{ \mathrm{#1} }}

\newcommand{\Dcal}{{\mathcal{D}}}

\newcommand{\Ncal}{{\mathcal{N}}}

\newcommand{\Rbb}{{\mathbb{R}}}

\newcommand{\Xbb}{{\mathbb{X}}}

\iftrue 
	\usepackage{xcolor}
	\hypersetup{
		colorlinks,
		linkcolor={blue!55!black},
		citecolor={red!55!black},
		urlcolor={blue!80!black}
	}
\fi
\newcommand{\Gp}{G_{\text{p}}} 
\newcommand{\Gv}{G_{\text{v}}} 
\newcommand{\Kp}{K_{\text{p}}} 
\newcommand{\Kv}{K_{\text{v}}} 
\newcommand{\Ti}{T_{\text{i}}} 
\newcommand{\Cp}{C_{\text{p}}} 
\newcommand{\Cv}{C_{\text{v}}} 

\newcommand{\Ts}{T_{\text{s}}} 

\newcommand{\CSS}{C_\text{SS}} 
\newcommand{\CSP}{C_\text{SP}} 
\newcommand{\CST}{C_\text{ST}} 
\newcommand{\Ccrit}{C_\text{crit}} 

\newcommand{\uk}{\underline{k}}
\newcommand{\ok}{\overline{k}}
\newcommand{\tSP}{t_\text{SP}} 
\newcommand{\tST}{t_\text{ST}} 
\newcommand{\kSP}{k_\text{SP}} 
\newcommand{\kST}{k_\text{ST}} 
\newcommand{\vcx}{\vc{x}}       
\newcommand{\nx}{n_{\vc{x}}}    

\newcommand{\GP}{{\mathcal{G}\!\mathcal{P}}}

\newcommand{\init}{{\text{init}}} 

\newcommand{\muf}{\mu^{\stiny{($f$)}}}
\newcommand{\mug}{\mu^{\stiny{($g$)}}}
\newcommand{\kf}{k^{\stiny{($f$)}}}
\newcommand{\kg}{k^{\stiny{($g$)}}}

\newcommand{\nuf}{\nu^{\stiny{($f$)}}}
\newcommand{\nug}{\nu^{\stiny{($g$)}}}
\newcommand{\vckf}{\vc{k}^{\stiny{($f$)}}}

\newcommand{\mxKf}{\mx{K}^{\stiny{($f$)}}}

\begin{document}

\title{Safety-Aware Cascade Controller Tuning Using Constrained Bayesian Optimization \thanks{ The authors are with Automatic Control Laboratory, ETH Zurich, and Inspire AG (e-mails: \{khosravm, rsmith, ralisa, lygeros\}@control.ee.ethz.ch, \{koenig,maier,rupenyan\}@inspire.ethz.ch).
This project has been funded by the Swiss Innovation Agency (Innosuisse), grant Nr 31695. Corresponding author A. Rupenyan rupenyan@inspire.ethz.ch}}
\author{Mohammad Khosravi, Christopher K\"onig, Markus Maier, Roy S. Smith, Alisa Rupenyan, and John Lygeros}
\maketitle

\setlength\tabcolsep{4pt}
\begin{abstract} 
This paper presents an automated, model-free, data-driven method for the safe tuning of PID cascade controller gains based on Bayesian optimization. The optimization objective is composed of data-driven performance metrics and modelled using Gaussian processes. We further introduce a data-driven constraint that captures the stability requirements from system data. Numerical evaluation shows that the proposed approach outperforms relay feedback autotuning and quickly converges to the global optimum, thanks to a tailored stopping criterion. We demonstrate the performance of the method in simulations and experiments. 
For experimental implementation, in addition to the introduced safety constraint, we integrate a method for automatic detection of the critical gains and extend the optimization objective with a penalty depending on the proximity of the current candidate points to the critical gains. The resulting automated tuning method optimizes system performance while ensuring stability and standardization.




\end{abstract}

 
\section{Introduction}

To replace the cumbersome and often suboptimal manual controller tuning several auto-tuning methods have been proposed in the literature over the years. A comprehensive review of tuning methods for PID controllers, including mechatronics applications, is provided in \cite{borase2020review}.
Established auto-tuning methods such as relay feedback tuning \cite{Astrom} can be applied to a cascade PID control structure \cite{Hang3}. Global parameter optimization algorithms, e. g. particle swarm \cite{Kennedy}, ant colony \cite{Dorigo} or genetic algorithms \cite{Mitchell}, previously used for tuning in \cite{Solihin}, \cite{Chiha} and \cite{Zhang2006}, might be optimal for the system performance, but require the availability of a precise model of the plant, or a vast number of trials to find the optimum. 

 The general idea  of optimizing performance indicators derived from system data for tuning of control parameters has been explored through various approaches, e.g., iterative feedforward tuning \cite{Li2019,IterTuning_Li2018}, variable gain selection \cite{VarGain_Li2015}, data-driven feedforward learning \cite{FeedforLearning_Song2020}, constrained optimization using interior point barrier algorithms \cite{Constrained_Radac2013}. The performance criteria can be represented by features in the data measured during system operation at different values of the controller parameters \cite{Khosravi2020,Khosravi2020b}.
 
Stability and safety are important both in tuning and in operation, especially in industrial systems. The tuning of a cascade control system can be formulated as a data-driven optimization problem, where the objective is to find the best controller parameters that fulfill both performance criteria, and safety constraints. A data-efficient method for optimizing a priori unknown objective function is Bayesian optimization (BO), where the unknown objective function is often modelled as a Gaussian process (GP). The GP model uses available data from prior evaluations to provide a mean and uncertainty estimate of the underlying function at new input locations. The BO then uses the resulting estimate for selecting new inputs, based on the optimization of an acquisition function, which combines the uncertainty or information content at the inputs, and the corresponding predicted performance. 

Optimization constraints can be modelled as separate GPs \cite{Gardner}, and the inputs corresponding to fulfilling the constraints with high probability are chosen by adapting the acquisition function. The approach has been demonstrated for manufacturing processes \cite{Maier, MaierGrind2020}, and for the run-to-run control of reluctance actuators \cite{BOcontrolRun2020}. Another approach to respect safety constraints, known as safe Bayesian optimization (SafeOpt)   provides probabilistic guarantees that all candidate samples remain in the constraint set \cite{Sui}. SafeOpt has been implemented and verified for robotic applications  \cite{Berkenkamp,Berkenkamp2}, and for controller tuning in heat pumps \cite{Khosravi} and shown to enable automatic and safe optimization of controller parameters. It has been further extended by an adaptive discretization based on particle swarms by \cite{Duivenvoorden} to efficiently perform BO in high-dimensional parameters spaces. 


 Controller tuning via BO is a suitable method to maximize the performance of industrial positioning systems by optimizing a combination of performance features \cite{Khosravi2020b}. However, when a system is embedded in a more complex equipment, additional forces might be applied in operation, or the system can be subject to wear, and the aspect of stability and safety gains priority. In this paper, we propose a safe, data-driven, model-free approach for the tuning of cascade controllers. The main contributions of the paper are: 1) We propose a safety-aware performance-based tuning algorithm, where safety and stability are ensured by informing the algorithm of the experimentally determined unsafe controller parameters; 2) The algorithm finds controller parameters with maximal performance; 3) We validate the approach numerically against benchmark safe tuning methods, and experimentally, on a linear axis component of a computer numerical control (CNC) grinding machine.



We first present the safety-aware tuning problem in section \ref{sec:PF}, then we describe the proposed approach for safe tuning in \ref{sec:BO}. The system modeling, performance metrics, and numerical implementation and verification of the algorithm are provided in section \ref{sec_model}, \ref{sec_BO}, and \ref{sec_simulation_1}, respectively. Section \ref{sec_experiments} presents the safety-aware tuning approach implemented on the experimental system.


\section{Safety-Aware Controller Tuning Problem}
\label{sec:PF} 
Consider a closed-loop system designed for controlling a possibly nonlinear plant.  Let $\vc{x}\in\Xbb$ be the vector of parameters describing the controller in the loop, where $\Xbb\subseteq\Rbb^{\nx}$ is the set of admissible controller parameters. These parameters should be designed such that the overall closed-loop performance is optimized and meanwhile the safety of the system is preserved. Based on the tracking error signal, i.e., the difference between the reference signal and the output signal, one can introduce various metrics indicating the quality of tracking, e.g., the magnitude of overshoot, the settling time, the steady-state error, the rising time, and other similar quantities. Given the reference signal and a fixed control structure, the error signal depends mainly on the vector of parameters $\vcx$. Accordingly, each of the above performance metrics is a function of $\vc{x}$, and hence, we denote them by $c_i:\Xbb\to\Rbb$, for $i=1,\ldots,n_\vc{c}$.
The overall performance metric, $f:\Xbb\to\Rbb$,  is defined as a weighted sum of the separate performance metrics, i.e., 
\begin{equation}\label{eqn:overall_perf}
f(\vcx):=\vc{w}^\tr \vc{c}(\vcx) = \sum_{i=1}^{n_{\vc{c}}}w_i c_i(\vcx),\qquad \vcx\in\Xbb,    
\end{equation}
where $\vc{c}(\vcx):=[c_i(\vcx)]_{i=1}^{n_{\vc{c}}}$ and $\vc{w}:=[w_i]_{i=1}^{n_{\vc{c}}}$.
The weights $w_1,\ldots,w_{n_{\vc{c}}}$ are decided based on the importance and the scale of the corresponding performance metric.
Similarly, one can consider a safety metric, $g:\Xbb\to\Rbb$, indicating the margin of the safety for a choice of controller parameters. For example, $g(\vcx)$ can be the overshoot observed in the response of the system.
Given the introduced metrics for performance and safety, we can introduce the 
\emph{safety-aware controller tuning problem}.
More precisely, to obtain the optimal controller gains $\vcx^*$ and meanwhile preserve the safety, we need to solve the following optimization problem
\begin{equation}\label{eqn:CBO}
\begin{array}{cl}    
\minOp_{x \in \Xbb}
&
f(\vcx)= 
\vc{w}^\tr \vc{c}(\vcx)
\\\text{s.t.}
&
g(\vcx) \leq g_{\text{tr}}\,, 
\end{array}
\end{equation}
where $g_{\text{tr}}$ is a safety threshold. 
For further safety improvement, one can consider additional performance metrics as barrier functions penalizing parameters close to critical gains.

\begin{figure}[t]
\centering
\includegraphics[width = 0.485\textwidth]{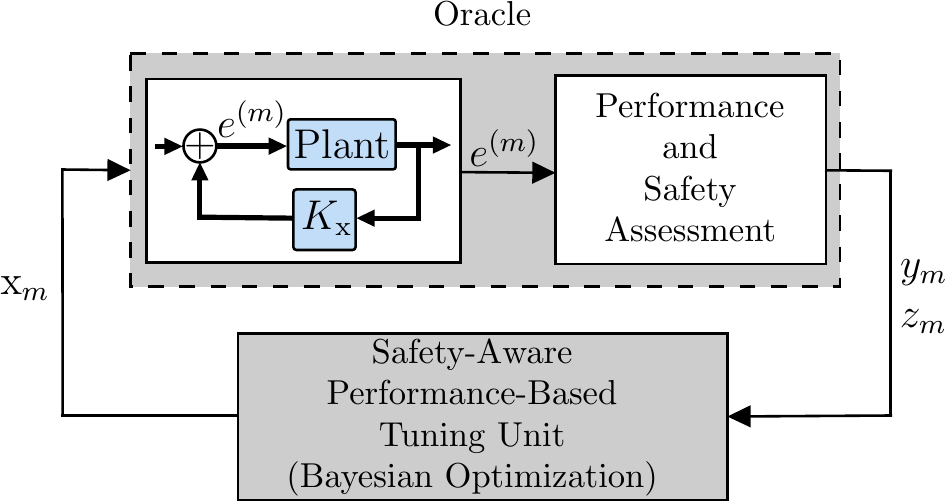}%
\caption{The safety-aware performance-based controller tuning scheme.}
\label{fig:scheme}
\end{figure}

The performance objective function $f$ and the safety constraint $g$ cannot be analytically calculated, i.e., they 
do not have a tractable closed-form expression, even when the dynamics of system are known. 
Indeed, with respect to each controller parameters, $\vc{x}$, 
one should perform an experiment on the system, measure the tracking error signal, and subsequently, calculate the value of $f(\vcx)$ and $g(\vcx)$.
In other words, each of these functions is given in a {\em black-box oracle} form.
Therefore, for tuning the controller gains via \eqref{eqn:CBO}, we utilize constrained Bayesian optimization (CBO), a data-driven approach for solving optimization problems whose objective function and constraints are available as expensive to evaluate black-box functions.
Accordingly, we have a data-driven tuning method which optimizes the closed-loop performance and preserves the safety of system (see Figure \ref{fig:scheme}).  
Note that this methodology can be extended to the control loop with multiple feedforward and feedback branches.

\section{Safe Controller Tuning Using Constrained Bayesian optimization}
\label{sec:BO}
The constrained Bayesian optimization (CBO) approach is based on sampling from the space of decision variables, $\Xbb$, and performing experiments. The scheme of controller tuning procedure based on this is shown in Figure  \ref{fig:scheme}. The detailed implementation is summarized in Algorithm \ref{alg:safety_aware_tuning} and explained in the following.

Assume that initially we are given a collection of controller gains  
$\Xbb_{\init}:= \{\vcx_i\in\Xbb\ \! |\ \! i=1,\ldots,m_\init\}$.
The algorithm maintains a pool of data $\Dcal_m:=\{(\vcx_i,y_i,z_i)\ \!|\ \! i=1,\ldots,m\}$,
where $\vcx_i$ are controller gains for which experiments have been performed, and, $y_i$ and $z_i$ are respectively the values of $f(\vcx_i)$ and $g(\vcx_i)$ computed from data collected in these experiments, i.e., the error signal. To capture uncertainty in the data collecting process we assume that $y_i=f(\vcx_i)+n_i$ and $z_i=g(\vcx_i)+\delta_i$, for all $i$,  where $n_i$ and $\delta_i$ are noise realisations. We assume that initially there are already $m\ge m_\init$ data points from the earlier operation of the machine and initialise $m=m_\init$. At each step of the algorithm 
we can perform a Gaussian process regression (GPR) for building surrogate functions for the performance metric and the constraint, denoted respectively by $f_m$ and $g_m$. 
More precisely, let $\GP(\muf,\kf)$ and $\GP(\mug,\kg)$ be Gaussian process priors respectively for $f$ and $g$, where $\muf:\Xbb\to\Rbb$ and $\kf:\Xbb\times\Xbb\to\Rbb$ are the mean and kernel functions in the GP prior taken for $f$, and similarly, $\mug:\Xbb\to\Rbb$ and $\kg:\Xbb\times\Xbb\to\Rbb$ are the mean and kernel functions in the GP prior taken for $g$.
Let the information matrices $\mx{X}_m$ and $\vc{y}_m$ be defined respectively as
$\mx{X}_m:=[\vcx_1,\ldots,\vcx_m]^\tr$, 
and 
$\vc{y}_m:=[y_1,\ldots,y_m]^\tr$. 
Then, we have $f_m(\vcx)\sim\Ncal\left(\muf_m(\vcx),\nuf_m(\vcx)\right)$, where the mean and variance are characterized as follows
\begin{align}
\label{eqn:post_fm}
\muf_{m}(\vcx) & := 
\muf(\vcx) \notag
\\&\quad\ \  + 
\vckf_{m}(\vc{x})^\tr 
(\mxKf_{m} + \sigma_{n}^{2}\eye)^{-1} 
(\vc{y}_m - \muf(\mx{X}_m)),  
\\ \label{eqn:cov}
\nuf_{m}(\vcx) &:=
\kf(\vcx,\vcx)
-
\vckf_{m}(\vc{x})^\tr  
(\mxKf_{m} + \sigma_{n}^{2}\eye)^{-1} 
\vckf_{m}(\vc{x}), 
\end{align}
where
the vector $\vckf_{m}(\vc{x}) = [\kf(\vcx,\vcx_1),\ldots,\kf(\vcx,\vcx_m)]^\tr$ , the vector
$\muf(\mx{X}_m) =
[\muf(\vcx_1),\ldots,\muf(\vcx_m)]^\tr,$
matrix  $\mxKf_{m}$ is defined as $[\kf(\vcx_i,\vcx_j)]_{i,j=1}^m$, $\eye$ is the identity matrix and $\sigma_n^2$ is the variance of the uncertainty in the evaluations of performance metric. 
Similarly, we have 
$g_m(\vcx)\sim\Ncal\left(\mug_m(\vcx),\nug_m(\vcx)\right)$
where the mean $\mug_m(\vcx)$ and variance $\nug_m(\vcx)$ are defined as in the previous case.

Given these probabilistic surrogate models for the performance metric and constraint, we can decide on the next candidate  sampling point, $\vcx_{m+1}\in\Xbb$, for the controller gains to be used in the next experiment. 
Using these GP models, one can consider an appropriate \emph{acquisition function} as the safe sampling criterion resulting in a suitable exploration-exploitation trade-off, (exploration of the regions of domain $\Xbb$ with the highest prediction uncertainty and exploitation of the points of $\Xbb$ with the lowest predicted cost), satisfying the safety constraint with high probability \cite{Frazier}.
 Let the \emph{expected improvement} function, $a_{\text{EI},m}:\Xbb\to \Rbb$, be defined as
\begin{equation}\label{eq:EI}
a_{\text{EI},m}(\vcx)
:=
\Big{(}\xi_m(\vcx)\Phi(\xi_m(\vcx))
+
\varphi(\xi_m(\vcx))\Big{)}\ \! \nuf_m(\vcx)^{\frac{1}{2}}, 
\end{equation}
where 
$\Phi$ and $\varphi$ are respectively the cumulative distribution function and probability density function of the standard normal distribution, and $\xi_m(\vcx):=(\muf_m(\vcx)-\vc{y}_m^+)/\nuf_m(\vcx)^{\frac{1}{2}}$ given that $\vc{y}_m^+$ is the best observed performance amongst the first $m$ experiments. 
Note that when the optimization problem is unconstrained, one can use $a_\text{EI,m}$ as the acquisition function  
aiming for a vector of controller gains with highest expected improvement of the performance metric $f$ \cite{Gardner}.
To include the constraint in \eqref{eqn:CBO}, one needs   to include the feasibility probability of the candidate sample. 
This results in 
 the \emph{constrained expected improvement}  \cite{Gardner},  defined as
\begin{equation}\label{eq:CEI}
a_\text{CEI,m}(\vcx)=  
\Phi\left(
\frac{g_{\text{tr}}-\mug_m(\vcx)}{\nug_m(\vcx)^{\frac{1}{2}}}\right)
a_\text{EI,m}(\vcx).
\end{equation} 
Subsequently, the next sampling point is obtained as 
\begin{equation}\label{eqn:x_next}
    \vcx_{m+1} := \argmax_{\vcx\in\Xbb}\ \! a_\text{CEI,m}(\vcx).   
\end{equation}
In order to solve optimization problem \eqref{eqn:x_next}, one can utilize various methods. 
Given $\vcx_{m+1}$, a new experiment is performed, the performance $y_{m+1}$ and constraint $z_{m+1}$ are evaluated, and the set of data is updated to $\Dcal_{m+1} = \Dcal_m \cup\{(\vcx_{m+1},y_m,z_m)\}$.
The sequence $\vcx_1,\vcx_2,\vcx_3,\ldots$ generated by this iterative procedure converges to the solution of \eqref{eqn:CBO} in a suitable sense \cite{Gardner}.

For the practical implementation of the controller tuning procedure, we need a stopping condition to ensure timely termination of the optimization algorithm. The stopping criterion can be based on the  maximal number of iterations, repeated sampling in the region of the minimal cost observed \cite{Khosravi2020b}, a threshold on the expected improvement acquisition function \cite{Nguyen}, or the convergence status of the uncertainty in the GP model of the cost \cite{Maier}. 
The stopping criterion used here depends on 
the ratio between the current expected improvement and the maximal expected improvement over all the previous iterations. 
This improves the robustness of the stopping condition with respect to the variation 
of the GP model hyperparameters. 
More precisely, the stopping condition is met when for three consecutive iterations we have
\begin{equation}
\label{eq:stop_aCEI}
a_\text{CEI}(\vcx_m) \leq \eta_\text{limit}\ \!\max_{0\leq i \leq m-1}a_\text{CEI}(\vcx_i),    
\end{equation}
where $\eta_\text{limit} \in [0,1)$
is a pre-determined threshold. 
The condition stops the iterative sampling procedure when the expected improvement does not change significantly by further exploration. 
The stopping iteration index is denoted by $m^*$.

The initial set of data $\Dcal_{m_{\init}}$ as well as the choice of kernel functions $\kf$ and $\kg$ play a significant role in the performance of the scheme.
Here to have maximally informative data set $\Dcal_{m_{\init}}$, initial sampling points $\Xbb_{\init}$ are drawn from a Latin hyper-cube experimental design \cite{mckay2000comparison}.
Subsequently, the hyperparameters of kernels $\kf,\kg$ 
are updated by minimizing the negative log marginal likelihood \cite{Rasmussen}, and kept fixed during the rest of the controller tuning procedure.
Though tuning the hyperparameters at every iteration might provide better GP models, it can disturb the convergence of the scheme.
While various options for the choice of kernel are introduced in the literature \cite{Rasmussen}; here 
we employ Matern kernels with parameter $\nu=\frac{3}{2}$ and automatic relevance determination, which enables the use of different kernel length scale parameters for each dimension of the input data \cite{Rasmussen}. 
Regarding the mean functions, we set $\muf=0$ and $\mug=0$ \cite{Rasmussen}. Note that one may employ other options 
when there is a prior knowledge on the structure of $f$ and $g$  \cite{Rasmussen}.


\begin{algorithm}[t]
\caption{Safety-Aware Controller Tuning Method}\label{alg:safety_aware_tuning}
\begin{algorithmic}[1]
\State \textbf{Input:} Set $\Xbb$, training data set $\Dcal_{m_\init}$, weight vector $\vc{w}$, GP priors $\GP(0,\kf)$ and $\GP(0,\kg)$, and  $\eta_\text{limit}$.  
\State Using $\Dcal_{m_\init}$ estimate the vectors of hyperparameters for $\GP(0,\kf)$ and $\GP(0,\kg)$, by minimizing the negative log marginal likelihood function.
\State $\Dcal_{m-1} = \Dcal_{m_\init}$ and $m-1= m_\init$.
\While{stopping condition is not met}
\State Derive $\vc{x}_{m}$ by solving \eqref{eqn:x_next}.
\State Set controller gains to $\vc{x}_{m+1}$, run the  experiment and measure the error signal $e^{(m)}$.
\State Obtain the values of the performance and the constraint functions $y_{m}$ and $z_{m}$ by calculating $c_i(\vcx)$, for $i=1,\ldots,n_{\vc{c}}$, and $g(\vcx)$.
\State Update the GP models: calculate $\muf_m,\nuf_m$ and $\mug_m,\nug_m$ (see \eqref{eqn:post_fm} and \eqref{eqn:cov}).
\State Update set of data: $\Dcal_m =  \Dcal_{m-1}\cup \{(\vcx_{m},y_{m},z_{m})\}$.
\EndWhile\label{DCP_loop}
\State \textbf{end} 
\State \textbf{Output:} $\vc{x}_{m^*}$.
\end{algorithmic}
\end{algorithm}

\section{Numerical Verification on a  CNC 
Machine}
\begin{figure}[h]
    \centering
    \includegraphics[width=0.45\textwidth]{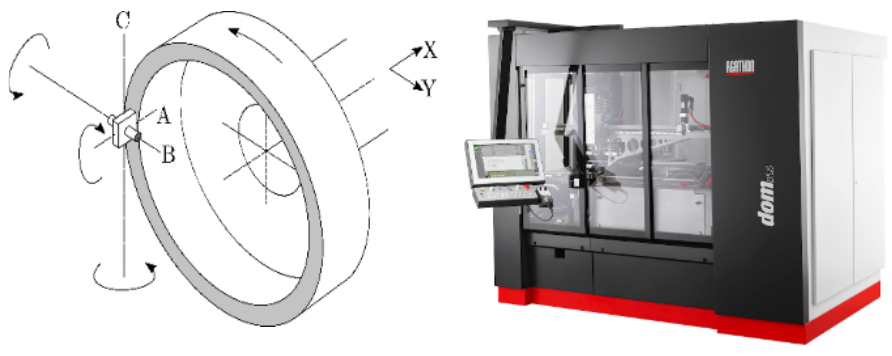}
    \caption{Left: The sketch of the linear $\mathrm{X, Y}$- and rotational $\mathrm{A, B, C}$- axis with respect to the grinding wheel. Right: The actual grinding machine.}
    \label{fig:machine_axes}
\end{figure}

The system of interest is a linear axis drive integrated in a four axis CNC grinding machine, endowed with two perpendicular linear axes moving the grinding wheel, and two rotational axes for the movement of a fixed work-piece  (See Figure \ref{fig:machine_axes}). 
Each axes is driven by a controlled brushless permanent magnet AC motor. 
The corresponding controllers are tuned independently in a special operational mode which does not involve grinding. 
The parameters of the controllers should be tuned to ensure adequate performance of the closed loop system. 
In this work, we are mainly focused on the linear axes $\mathrm{X}$ and $\mathrm{Y}$ for moving the grinding wheel in the direction perpendicular to the ground surface,  due to its critical impact on the grinding quality. We report the tuning of the $\mathrm{X}$-axis controller, although we implemented the method also for the $\mathrm{Y}$-  and for the rotational $\mathrm{B}$- axis with similar results.
The same method can be applied to other axes in the grinding machine.

\subsection{System Structure and Modelling}
\label{sec_model}

\begin{figure}[h]
\centering
\includegraphics[width=0.485\textwidth]{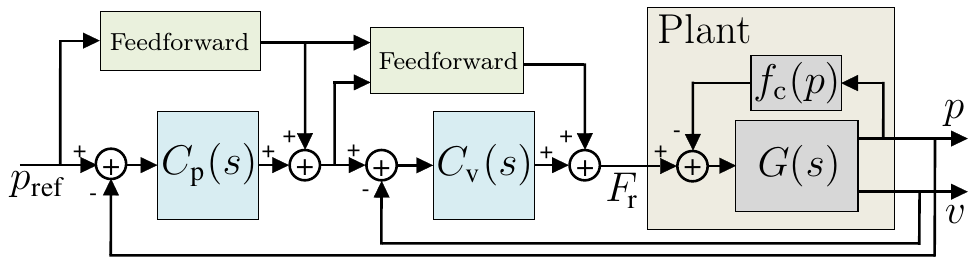}
\caption{The plant and the control structure of the linear axis.}
\label{fig:Control_Structure}
\end{figure}

The plant is modeled as a combination of a linear sub-system and a nonlinear sub-system, as shown in Figure \ref{fig:Control_Structure}. 
The linear block models the linear axis as a damped single mass system, following \cite{Khosravi2020}, 
with transfer function $G(s)$ defined as 
$G(s) = \begin{bmatrix}
\Gp(s)&\Gv(s)
\end{bmatrix}^\tr$,
where $\Gp(s)$ and $\Gv(s)$ are the transfer functions respectively from force to position and from force to velocity. 
These transfer functions mainly depend on the mass of the axis, $m$, and also the damping coefficient $b$ due to the friction. 
These values are estimated using a least squares fitting identification procedure and provided in Table \ref{tab:ForceRipple}. 
The nonlinear subsystem models the nonlinear phenomena of force ripple and cogging effects in the permanent magnet motor.
Force ripple is caused by the irregularity of the magnetic field of the permanent magnet and the inaccuracy of the electronic communication by the servo amplifiers \cite{Rohrig}, while the cogging force results from the attraction between the ferromagnetic core of the motor windings and the permanent magnets on the rail \cite{Villegas}. 
These forces depend on the position with an almost periodic behaviour. Following \cite{Rohrig}, we model them by Fourier truncated expansion:
\begin{equation}\label{eq:Force_Ripple_new}
f_\vc{c}(p) = c_1 + c_2p +  
\sum_{k=1}^{n} c_{2 k+2}\!\ \sin\!\Big{(}\frac{1}{c_{3}}2 k \pi p+c_{2 k+3}\Big{)}\, , \end{equation}
where $p$ is the position, $c_1$ is the average thrust force, $c_2$ is the gradient of the curve caused by sealing bellows, $c_3$ is the largest dominant period  described by the distance of a pair of magnets in the rail, $n$ is the number of modelled frequencies, and, $c_{2k+2}$ and $c_{2k+3}$ are respectively the amplitudes and the phase shifts of the sinusoidal functions, for $k=1, 2,\ldots,n$. 
The vector of parameters 
$\vc{c}:=[c_1,\ldots,c_{2n+3}]^\tr$
are estimated using least squares error minimization between the modelled ripple and cogging forces in \eqref{eq:Force_Ripple_new} and the measured force signal at constant, non-zero velocity (to cancel the effects from linear dynamics). The estimated values of the parameters are provided in Table \ref{tab:ForceRipple}.  In our case, a single harmonic suffices to capture most of the nonlinearity present in the system.
\begin{table}[b]
 \caption{Parameters of the system}
 \label{tab:ForceRipple}
    \centering
    \ra{1.1}
    \begin{tabular}{ @{}l l l @{} }
    \toprule
     Parameter & Value & Unit\\
     \midrule
    $n$ number of frequencies & $1$ &-\\
    $c_1$ average thrust force & -$104.9$& N\\
    $c_2$ gradient & $682.44$ & N/m\\
    $c_3$ distance of pair of magnets& $2.364\text{e-}1$&m\\
    $c_4$ amplitude & $23.55$&N\\
    $c_5$  phase shift& $8.77\text{e-}7$&m\\ 
    $m$ axis mass & $388.61$ & kg\\
    $b$ damping coefficient  & $2224.60$ &kg/s\\
    \bottomrule
    \end{tabular}
\end{table} 

The linear axis is controlled by a three-level cascade controller shown in Figure \ref{fig:Control_Structure}. 
The outermost loop controls the position by P-controller $\Cp(s)=\Kp$, and the middle loop controls the velocity by PI-controller $\Cv(s) = \Kv(1+\frac{1}{\Ti s})$. 
The innermost loop of the control structure is a current controller for the permanent magnet AC motor of the linear drive, which is pre-tuned robustly and hence treated as a part of the plant $G(s)$.  
As illustrated in Figure \ref{fig:Control_Structure},
feedforward structures are used to accelerate the response of the system. 
Similar to the current controller, the gains in these feedforward blocks are set initially and not modified during the tuning procedure.
Accordingly, in the followings, we consider two cases for the controller parameters to be tuned:  $\mathrm{x} = (\Kp, \Kv, \Ti)$ and $\mathrm{x} = (\Kp, \Kv)$, where in the latter one $\Ti)$ is set to $\Ti= 7.5$ms.
\subsection{Safe Controller Tuning for the CNC Machine}
\label{sec_BO}

During standard operation, the grinding machine moves back and forth, starting from a reset point, moving towards the operation location, remaining there to perform the grinding mission, and then returning to the setback point.
Accordingly, the position reference for the closed-loop system has a trapezoidal shape.
Our aim is to tune the controller gains, $\mathrm{x}$, such that the vibration during grinding and the amount of overshoot at arriving or leaving the operation point should be minimized, while the closed-loop system safety is guaranteed (see Section \ref{sec:PF}).
To this end, based on the closed-loop behaviour of the system, we consider performance metrics illustrated in Figure \ref{fig:cost}.
More precisely, let $\Ts$ denote the sampling time and $e_{\vcx}:[\uk\Ts,\ok\Ts]\to\Rbb$ be the error signal during a cycle of standard operation of the system where the controller gains are set to $\vcx$.
Also, let $\tSP:=\kSP\Ts$ and $\tST:=\kST\Ts$ be  respectively the time instants of arriving or departing the operation location.  
%
Accordingly, the performance metrics are defined as following 
\begin{itemize}
    \item $\CSP(\vcx)$, the maximum error after departing the operation location, i.e., $\CSP(\vcx) := \max_{\kST+1 \le k}\ |e_{\vcx}(k\Ts)|$,
     \item $\CST(\vcx)$, the overshoot after arriving at operation location,
     i.e., $\CST(\vcx) := \max_{\kSP \le k \le \kST}\ |e_{\vcx}(k\Ts)|$,
     \item$\CSS(\vcx)$, the $L_1$-norm of the position error at the operation location scaled with the sampling time $\Ts$,
     i.e., $\CSS(\vcx) := \Ts \sum_{\kSP \le k \le \kST}\ |e_{\vcx}(k\Ts)|$,
     \item{$C_\text{crit}(\vcx)$, a metric introduced for further enhancing experimental safety (see equation \eqref{eqn:Ccrit}). 
     }
\end{itemize}
\begin{figure}[b]
\centering
\includegraphics[width=0.4\textwidth]{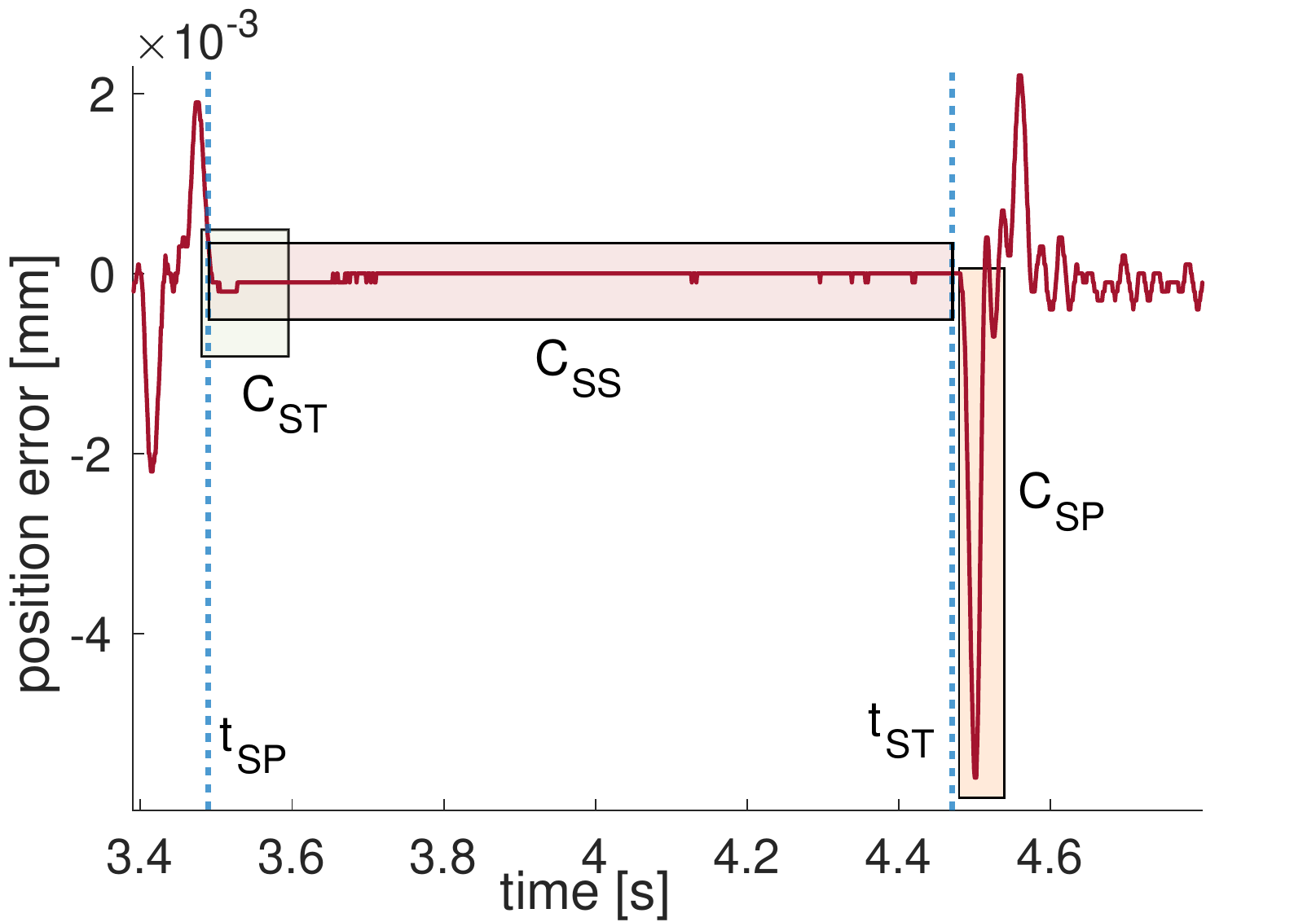}
\caption{Illustration of the three performance metrics on the position error signal (input signal to $C_{\mathrm{p}}(s)$)}
\label{fig:cost}
\end{figure}
Due to static friction,  $\CSP(\vcx)$ has a larger magnitude compared to $\CST(\vcx)$  (see Figure \ref{fig:cost}), however, 
$\CST(\vcx)$ is more sensitive to the instabilities of the system.
Therefore, we additionally consider a constraint $\CST(\vcx) \leq c_{\text{b}}$ for enforcing the safety, where $c_{\text{b}}$ is a pre-decided bound for the maximum allowed overshoot.
Accordingly, we have
\begin{equation}
\vc{c}(\vcx):=[\CSP(\vcx), \CSS(\vcx),\CST(\vcx),\Ccrit(\vcx)]^\tr,
\end{equation}
and the overall performance metric $f(\vcx)$ is defined as the weighted sum introduced in \eqref{eqn:overall_perf}.
From Section \ref{sec:PF}, the optimal controller gains $\vcx^*$ are obtained by solving the following optimization problem
\begin{equation}\label{eqn:CBO_NumVer}
\begin{array}{cl}    
\minOp_{x \in \Xbb}
&
\!\!
f(\vcx):= 
w_1 \CSP(\vcx) \!+\! w_2 \CSS(\vcx) \!+\!  w_3 \CST(\vcx) \!+\!  w_4 \Ccrit(\vcx) 
\\\text{s.t.}
&
\!\!
g(\vcx):=\CST(\mathrm{x}) \leq c_{\text{b}}.
\end{array}
\end{equation}
One can see that functions $f$ and $g$ are in black-box oracle form with respect to $\vc{x}$, i.e., their analytical closed form is not available and one can access their values for a given $\vc{x}$ only by performing an experiment on the machine and a subsequent calculation.
Accordingly, for tuning the controller gains, we utilize the data-driven procedure introduced in Section \ref{sec:BO}. For the numerical experiments, it was sufficient to maintain safety only through the constraint $g(\vcx)$, therefore the weight $w_4$ corresponding to $\Ccrit$ was different than zero only for the experiments on machine.

The function $C_\text{crit}(\mathbf{x})$ is defined as
\begin{equation}\label{eqn:Ccrit}
C_\text{crit}(\vcx) = \rho_\text{crit}\ \! \exp(\frac{\Kp}{\Kp^\text{crit}})\ \! \exp(\frac{\Kv}{\Kv^\text{crit}}) \, ,
\end{equation}
where $\rho_\text{crit}$ is a scaling factor to adjust the magnitude of the penalty according to the other terms of the cost,  $\Kp^\text{crit}$ and $\Kv^\text{crit}$ are the controller critical gains estimated experimentally during training data collection. $C_\text{crit}(\vcx)$ is a penalization term introduced to incorporate additional information on safety. 

\subsection{Numerical Results}
\label{sec_simulation}


In this section, we numerically evaluate the performance of the proposed tuning method and compare with competing methods in the literature.
\subsubsection{Global Optimum and Bayesian Optimization Range}
\label{sec_simulation_1}
First, we use the linear axis model from Section \ref{sec_model} to compute the optimization ranges where the system is stable, and to find the global optimum using grid search. The ranges are later used in the experimental implementation.

\begin{figure}[h]
\centering
\includegraphics[width=0.485\textwidth]{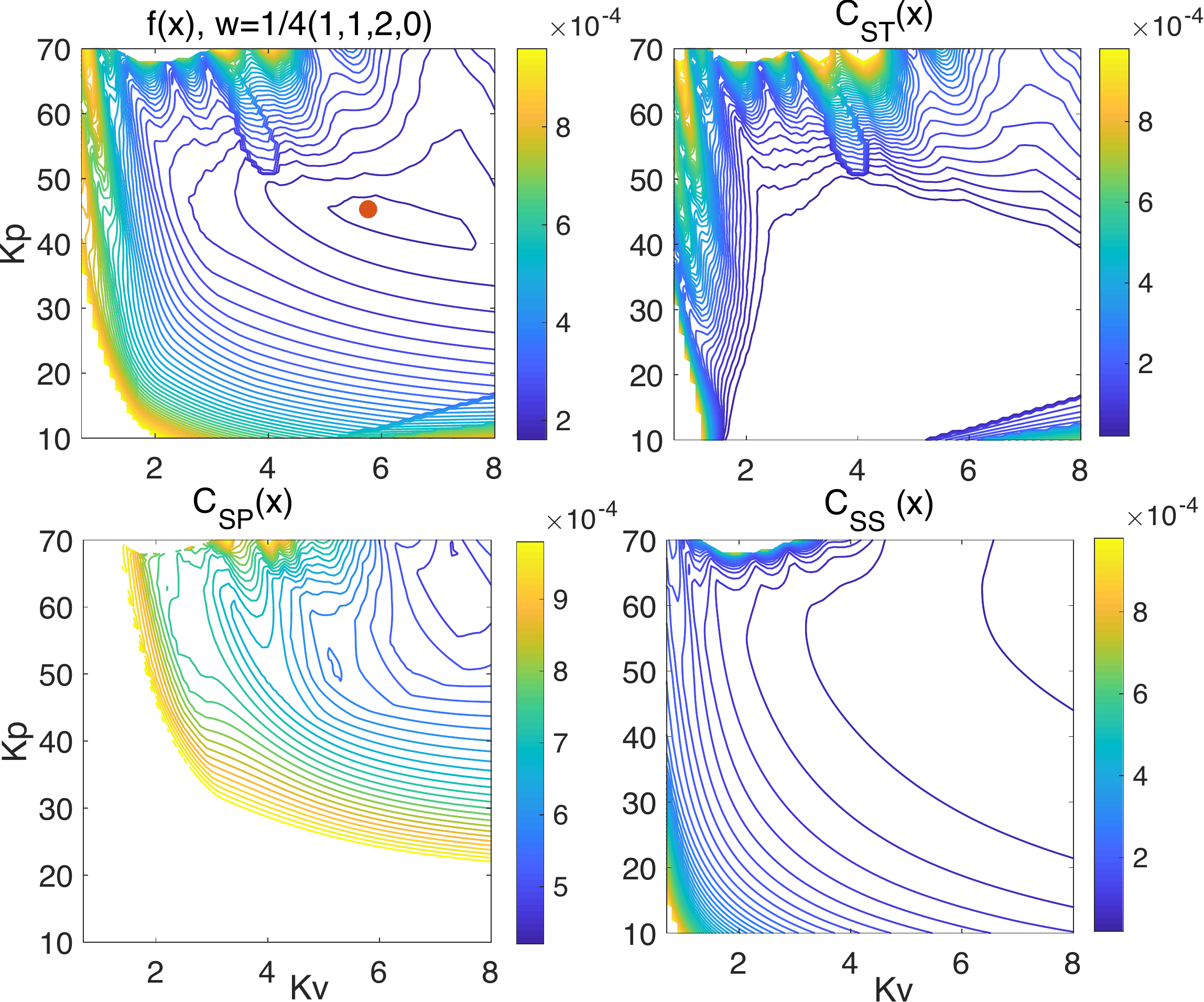}
\caption{2D grid of the individual and combined cost terms using the identified system model, $\Ti = 7.5$ms. The optimum resulting from grid evaluation is shown in the top left panel.}
\label{fig:GX_sim_grid_7p5ms}
\end{figure}

In Figure \ref{fig:GX_sim_grid_7p5ms}, the individual cost terms and their weighted sum $f(\vcx)$ are calculated and displayed on a two-dimensional grid $\Kp \times \Kv$ with 320 $\times$ 300 points and resolution $(\Delta \Kp,\Delta \Kv) = (0.25, 0.05)$, for fixed $\Ti = 7.5$ms. 
The computed global optimum is shown on the upper left cost panel. 
Table \ref{tab:opt_range} shows the parameter values corresponding to the computed optima for $\Kp$ and $\Kv$, while keeping $\Ti$ fixed, and those corresponding to joint exhaustive evaluation on the grid of the three parameters where $\Ti$ is a variable as well. 
The latter finds a solution with lower cost, while the values of the parameters are increased.
The range of the parameters' values chosen here 
(see Table \ref{tab:opt_range})
includes the nominal and optimal parameters as well as the stability limits resulting from grid evaluation, where the overshoot criterion is violated.
\begin{table} [h]
 \caption{Global optima and optimization range for the numerical exhaustive evaluation experiments}
 \label{tab:opt_range}
    \centering
    \ra{1.3}
    \begin{tabular}{ @{}l c c c c c c@{} }\toprule
    & 2D optimum && 3D optimum && Range\\
    \midrule
     $\Kp$ [$\mathrm{\frac{1000}{min}}$] & $45.5$ && $57$  &&  $10$ -- $70$\\
     $\Kv$ [$\mathrm{\frac{N}{mm/min}}$] & $5.9$ && $6.85$ && $0.5$ -- $8$\\
     $\Ti$ [ms] & $7.5$ (fixed) && $12.5$ && $5$ -- $17$\\
     Cost $f$ & $1.5048$e-$4$ && $1.3923$e-$4$ && --\\
   \bottomrule
    \end{tabular}
\end{table}

\subsubsection{Comparison with Relay Feedback Tuning}
\label{sec_simulation_4}
For the next numerical experiment we compare the proposed method with relay feedback tuning which is a benchmark controller (auto)tuning approach 
\cite{Hang3}. The relay feedback tuning is fully deterministic, robust to the disturbances, and tunes simultaneously the three parameters of the position and velocity controller $\Kp$, $\Kv$ and $\Ti$. 
The tuning procedure depends on information automatically extracted from the frequency response of the plant, which is generally sufficient for PID controller tuning of many processes \cite{Hang2,Hang3}. 
One may utilize the relay method for nonlinear processes when it is possible to tune the relay parameters to keep the system in a region with approximately linear behavior.

Table \ref{tab:RF_comp} shows the results of ten simulated experiments with different training samples drawn using Latin hypercube sampling for the joint optimization of $\Kp$, $\Kv$ and $\Ti$ in comparison with relay feedback tuning. We use particle swarm optimization to optimize the acquisition function.
\begin{table} [t]
\caption{Comparison of the parameters, cost and number of iterations between the proposed method with $\vc{w}=\frac{1}{4}(1,1,2,0)^\tr$ and relay feedback tuning (mean and 95\% confidence interval
)}
\label{tab:RF_comp}
    \centering
    \ra{1.3}
    \begin{tabular}{ @{}l c c c c@{} }\toprule
   Parameter & CBO && Relay feedback\\ \midrule
     $\Kp$ [$\mathrm{\frac{1000}{min}}$] & $57.64\pm5.67$ && $41.72$\\
     $\Kv$ [$\mathrm{\frac{N}{mm/min}}$] & $6.48\pm 0.77$ && $3.64$\\
     $\Ti$ [ms] & $13.90\pm 0.32$ && $16.80$\\
     Cost $f$ & $1.48$e-$4 \pm 0.11$e-$4$ && $2.78$e-$4$\\
     Number of iterations $N$ & $71 \pm 31$ && $1$\\
   \bottomrule
    \end{tabular}
\end{table}
Since the proposed tuning method is performance-based, the cost is 48\% lower than the corresponding cost of the parameters tuned by the relay feedback procedure. 
However, the algorithm requires between 40 and 100 iterations in addition to the 25 evaluations of the initial set, while the relay feedback method calculates the parameters from a single response of the system \cite{Khosravi2020,Khosravi2020b}. 
Accordingly, one can leverage the benefit of the proposed method especially when the machine performs a repetitive task.
Both methods can be fully automated, and could be selected depending on the performance of the system and the duration of the tuning procedure.




\subsubsection{Safety and Stability}
\label{sec_simulation_3}
Avoiding unstable parameter values, and keeping the system safe is critical both during tuning and operation. 
We explored tuning using an adaptation of SafeOpt, a BO-based algorithm for safe exploration of the parameter space \cite{Duivenvoorden}. The SafeOpt algorithm starts with an initial safe set and uses two different acquisition functions for sampling. The first one is defined based on the lower confidence bounds obtained from Gaussian process model of the objective function, and it is designed to find candidate optimizers. 
The second acquisition function is for exploring the domain and expanding the safe set of the parameters. This acquisition function is defined based on the confidence bounds obtained for the safety constraint, and penalizes candidate points which are probably unsafe. Hence, the input parameters fulfilling the safety constraints are selected with high certainty. 
\begin{figure}[h]
\centering
\includegraphics[width=0.42\textwidth]{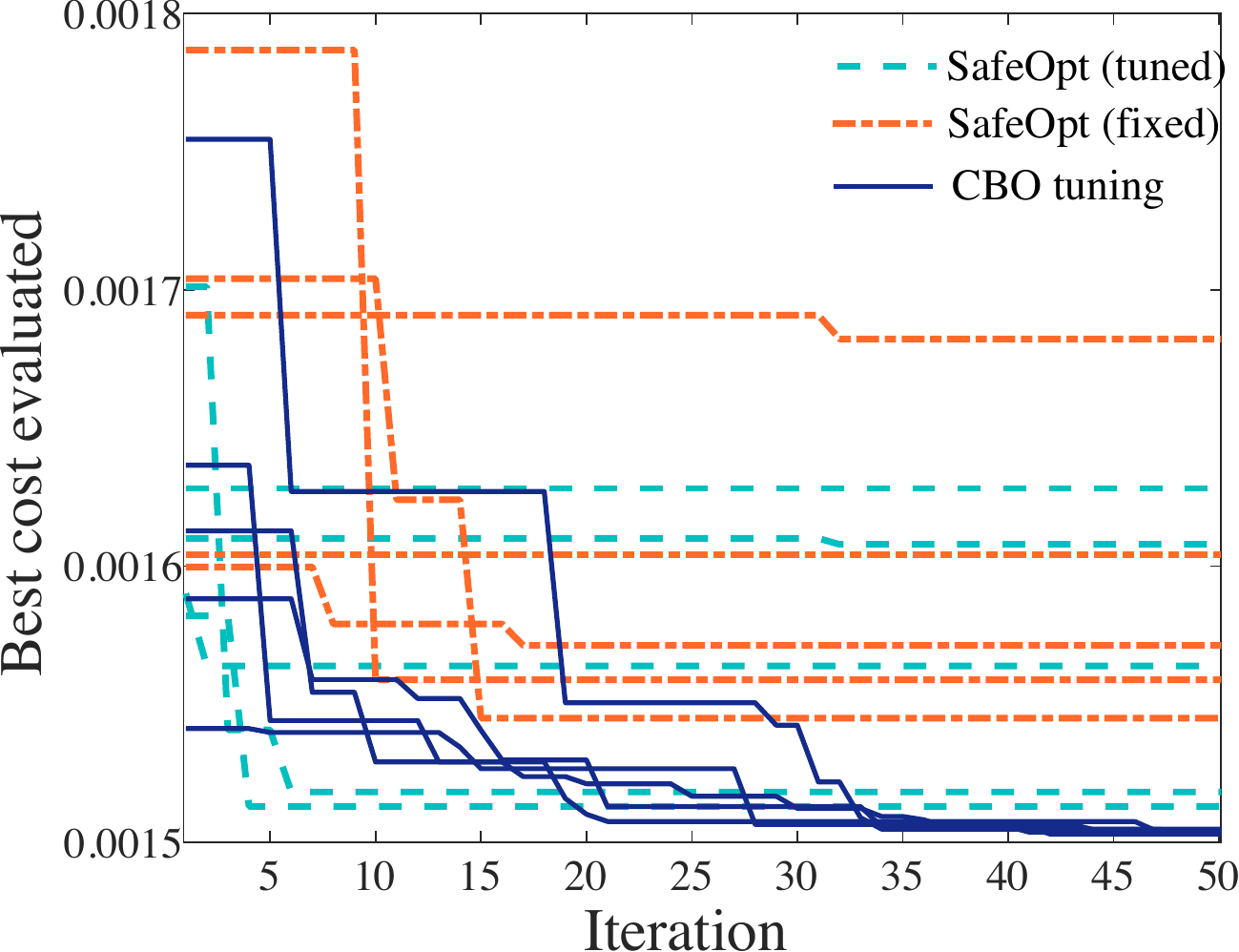}
\caption{Comparison of the convergence of the CBO algorithm with SafeOpt}
\label{fig:GX_sim_SafeOpt_comp}
\end{figure}
Figure \ref{fig:GX_sim_SafeOpt_comp} shows the performance achieved by SafeOpt and by constrained Bayesian optimization using $f(\vcx)$ for the optimization of $\Kv$ and $\Kp$. 
A training set of 15 measurements is selected by Latin hypercube sampling inside the range $20 \leq \Kp \leq 52.5$ and $1 \leq \Kv \leq 6$, computed using grid evaluation, to avoid areas that violate the constraint. 
The SafeOpt particle swarm implementation uses the samples of the initial set as safe starters. 
The hyperparameters of both algorithms are tuned by minimization of the negative log marginal likelihood, after evaluation of the initial set. 
In a second series of refined SafeOpt experiments, the hyperparameters of the constraint are fixed to more conservative values, namely smaller length-scales and higher signal variance, resulting in higher uncertainty outside known regions.

Each optimization experiment terminates after running a fixed number of 50 iterations. At the end of the 50 iterations both implementations of the SafeOpt algorithm, including experiments that start with better values inside the training set, remain with higher cost as compared to CBO.
SafeOpt needs to complete PSO three times for every iteration (see \cite{Duivenvoorden} for more details), which increases the computational time. 
Due to the large distance between the optimum and the constraint, SafeOpt with tuned hyperparameters results in more constraint violations than the CBO algorithm (see Table \ref{tab:SafeOpt_comp}), because SafeOpt actively tries to expand the safe optimization range of the function and thus also chooses samples near the boundary of feasible set. 
The constraint violations can be reduced 
by fixing the hyperparameters, which leads to slower convergence. 
We have thus adopted CBO in combination with additional safety measures for the real system.
\begin{table} [t]
\caption{Comparison of CBO with SafeOpt for the optimization of $\Kp$ and $\Kv$ (mean and 95\% confidence interval).}
\label{tab:SafeOpt_comp}
    \centering
    \ra{1.3}
    \begin{tabular}{ @{}l c c c c@{} }\toprule
     & & Median & Median\\    
     & Final cost & comp. time & constr. violations\\ \midrule
     CBO & 1.50e-4 $\pm$ 5.59e-7 & 303.5 & 1\\
     SafeOpt (tuned) & 1.55e-4 $\pm$ 2.08e-6 & 710.2 & 5\\
     SafeOpt (fixed) & 1.59e-4 $\pm$ 1.28e-5 & 707.3 & 0\\
   \bottomrule
    \end{tabular}
\end{table}

\section{Experimental Results}
\label{sec_experiments}

We now demonstrate the tuning method on the linear axis drive. First, we introduce the system where we do experimental validation. We then present the adaptation needed to safely operate and reach stable parameters. Finally, we examine the convergence of the algorithm.

\subsection{Experimental System}
\label{sec:exper_system}
Our experimental system is a linear axis drive integrated in a DomSemi grinding machine (see Figure \ref{fig:machine_axes}) for face grinding from Agathon AG, steered by a controlled brushless permanent magnet AC motor.
The position and velocity measurements are provided with an optical sensor system with precision of $100$nm.
The nominal controller settings used in the model and in the real system for the position P and velocity PI controller are ($\Kp^\text{nom}=20~[\mathrm{1000/min}]$, $\Kv^\text{nom}1~[\mathrm{N /(mm/min)}]$, $\Ti^\text{nom} = 7.5~[\text{ms}]$).
The position, velocity and force of the linear axis drive are limited by the controller to a $137$mm range, $315$mm/s speed and $400$\% of the motor power in both directions, respectively. 

\subsection{Critical Gain Detection}
\label{sec_experiments_1}

\begin{figure}[t]
\centering
\includegraphics[width=0.485\textwidth]{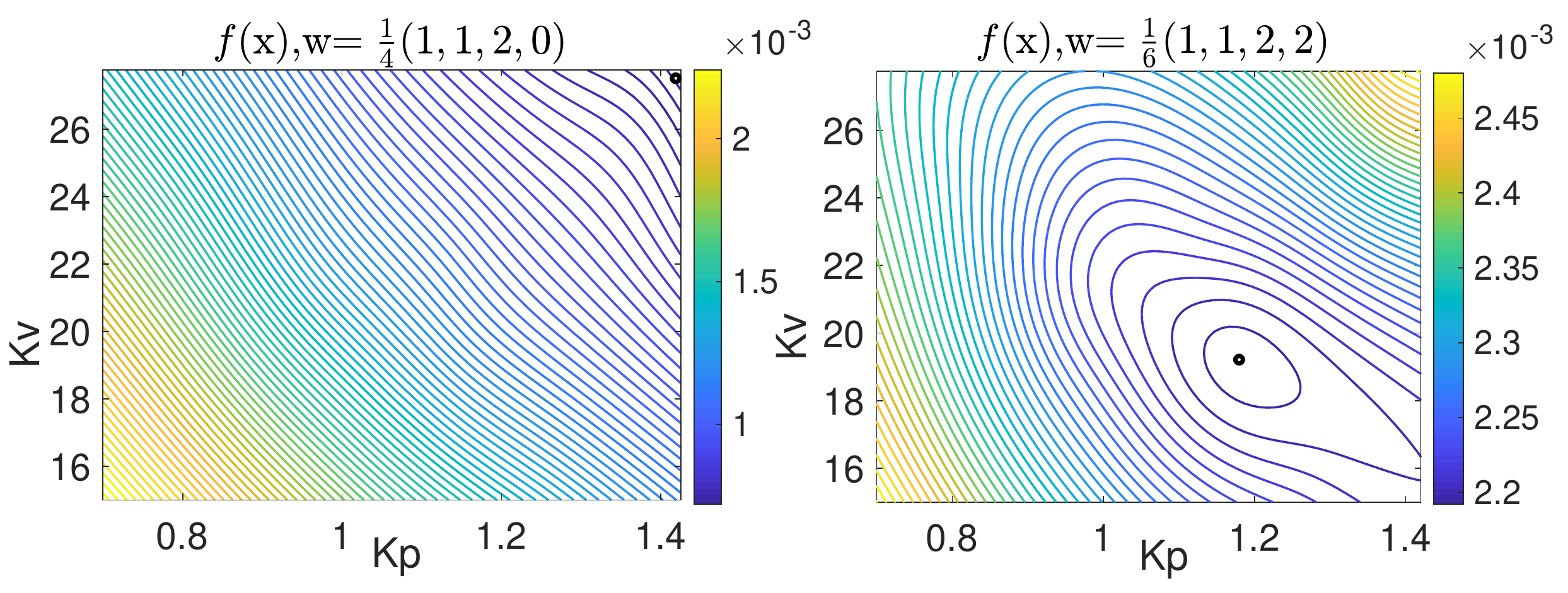}
\caption{Predicted cost for tuning on the actual axis with and without penalty. The optima are shown by black circles.}
\label{fig:GX_prediction_exp}
\end{figure}

We implement the tuning method using the optimization as formulated in \ref{eqn:CBO} to tune $\Kp$ and $\Kv$. The corresponding weights in the cost are $\vc{w}:=\frac{1}{4}(1,1,2,0)^\tr$. We restrict the tuning on the real system in the range $20 \leq \Kp \leq 52.5$ and $1 \leq \Kv \leq 6$, computed using our numerical model with grid evaluation. The left panel of Figure \ref{fig:GX_prediction_exp} shows the resulting GP model of the cost after completing all tuning iterations, along with the corresponding optimized controller parameters. The optimum $(\Kp,\Kv) = (27.75,1.43)$ is located right at the limit of the safe range, as can be seen on the left panel of Figure \ref{fig:GX_prediction_exp}. While this is acceptable in the tuning mode, during actual grinding the performance can shift due to additional loads and forces, and the optimum could accordingly become unsafe. Routinely this is overcome by selecting conservative controller parameters, which in turn curbs the system's performance.

\begin{figure}[h]
\centering
\includegraphics[width=0.485\textwidth]{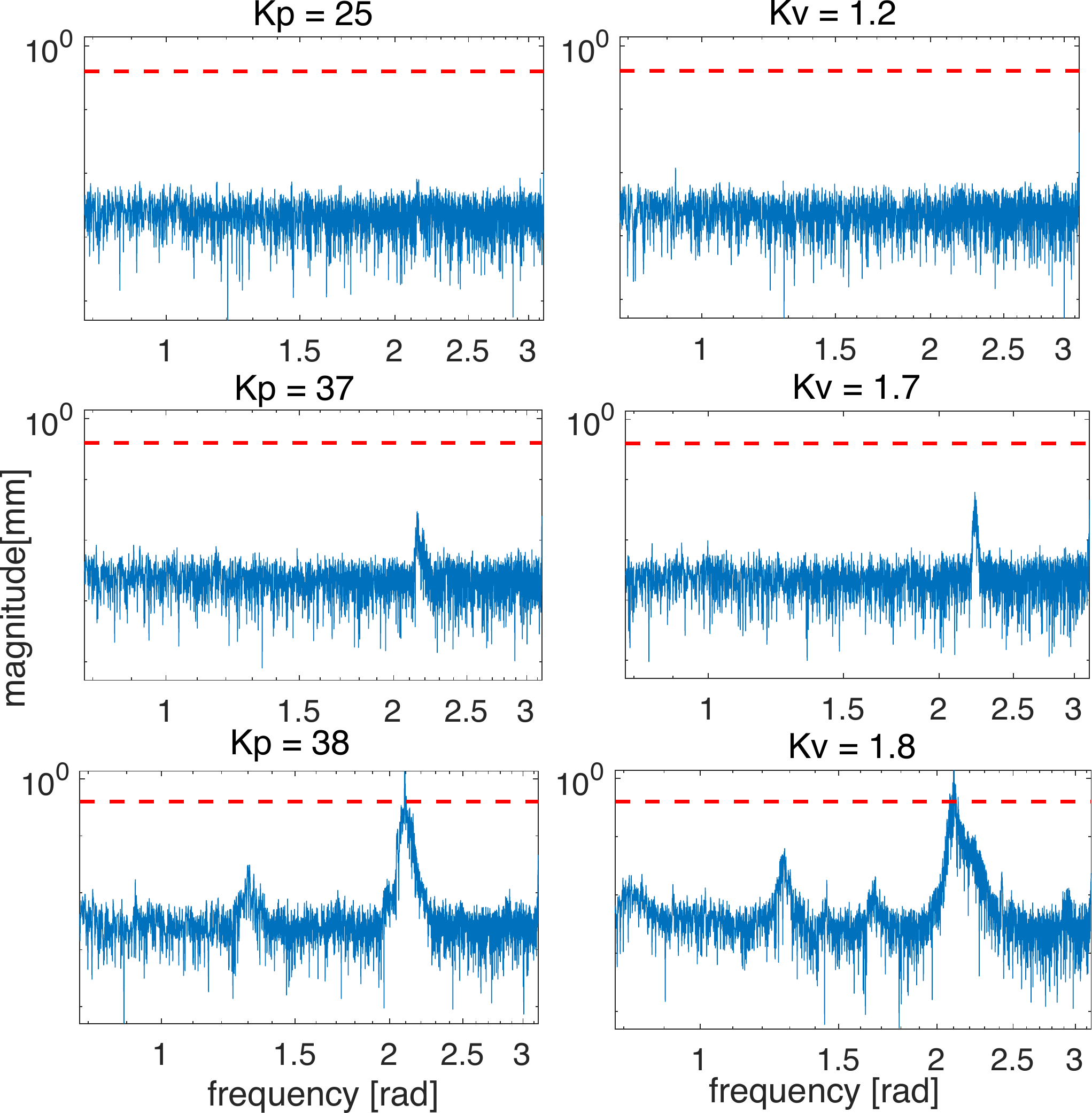}
\caption{FFT of the position error signal for different controller gains, with the corresponding threshold indicated with a dashed line. The bottom panels are examples of reaching the critical gain, following consecutive scanning for $\Kv$ and $\Kp$.}
\label{fig:GX_detcrit}
\end{figure}

We incorporate this additional uncertainty by increasing the weight for the safety penalty $C_{\mathrm{crit}}$ from $0$ to $1/3$ and by adapting the remaining weights. We determine the critical gains in the $C_{\mathrm{crit}}$ by monitoring the fast Fourier transform (FFT) of the position error in a fixed frequency window. This comprises an initial scan for all gain parameters, starting from the nominal values and increasing until a threshold of $0.4$mm in the FFT of the position error is exceeded, as shown in Figure \ref{fig:GX_detcrit}. The critical gains in each loop are determined in consecutive scans, while keeping the parameters of the inner loop fixed, therefore discarding the mutual dependence of the controller's parameters. 
Once these gains are known, the cost penalization in \eqref{eqn:CBO} is applied. The right panel of Figure \ref{fig:GX_prediction_exp} shows the 
GP model predictions of the cost $f(\vcx)$ with 
weights
$\vc{w}:=\frac{1}{6}(1,1,2,2)^\tr$, after collecting the data from all BO iterations on the real system, along with the optimized controller gains. 
In this case, the optimization range  is limited and depends on the detected critical gains in the scanning phase, i.e., $\Kp\le 0.75 \ \! \Kp^\text{crit}$ and $\Kv\le 0.75 \ \! \Kv^\text{crit}$. This is the range used for both panels in Figure \ref{fig:GX_prediction_exp}.
\subsection{Convergence of Method}
\label{sec_experiments_2}
\begin{figure}[h]
\centering
\includegraphics[width=0.4\textwidth]{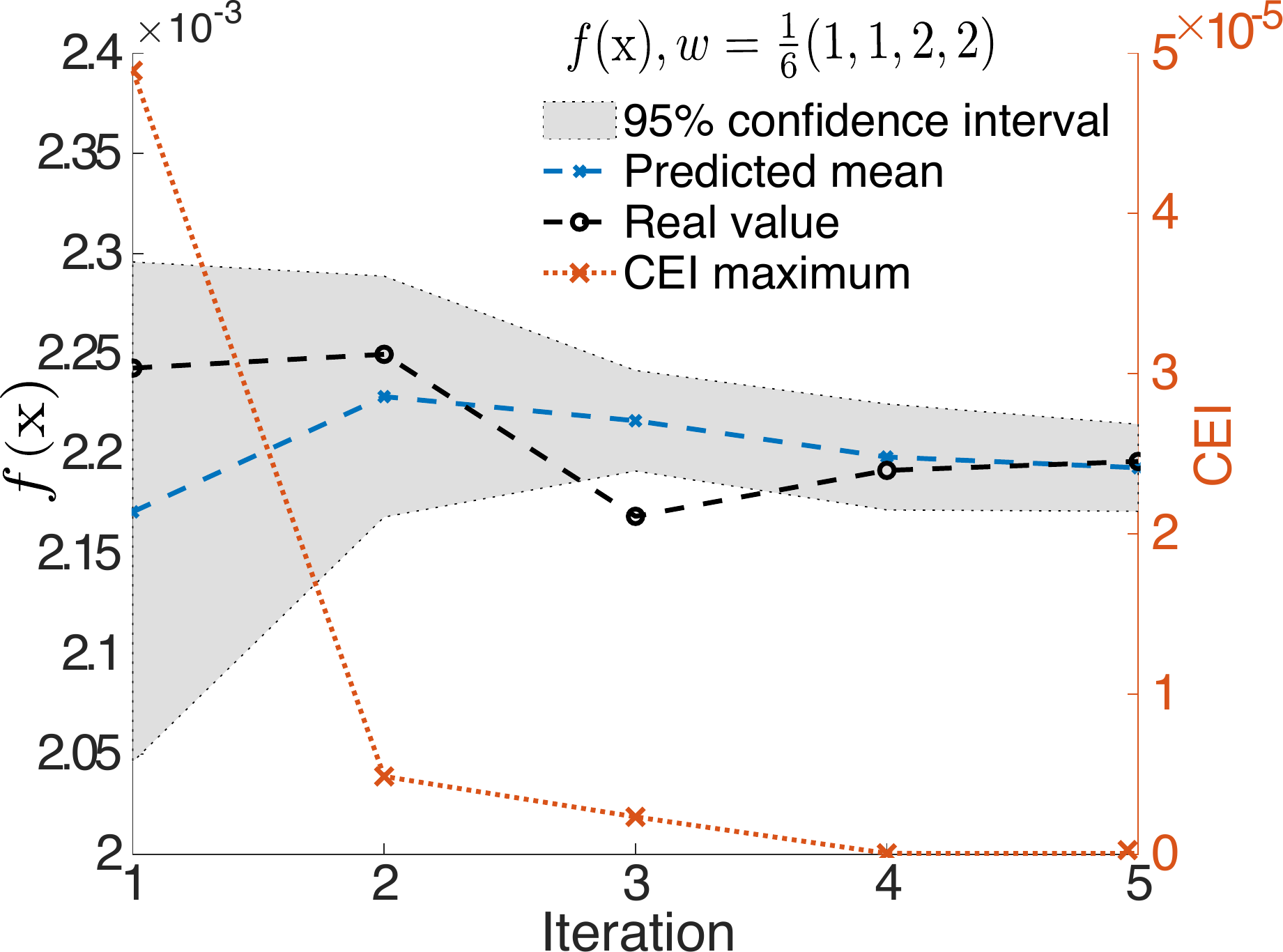}
\caption{The predicted cost, real cost and maximum CEI over all iterations of the proposed tuning algorithm on the real grinding system.}
\label{fig:GX_experimentBO_meancon}
\end{figure}
The convergence of the constrained expected improvement (CEI) maximum and the variance of the CEI maximum candidate cost (the predicted mean of the cost) are illustrated in Figure \ref{fig:GX_experimentBO_meancon} for each iteration of the proposed tuning scheme, and the acquisition function evolution is included as well. 
The predicted mean of the cost decreases with the number of iterations, and already at the fourth iteration is virtually identical to the experimentally determined cost. 
After the CEI maximum of the current iteration drops under 5\% of its maximal value in iteration 3, which is the threshold set in the stopping criterion \eqref{eq:stop_aCEI}, and stays there for 3 consecutive iterations, the algorithm is terminated after iteration 5. Although the minimum of the cost happens at iteration 3, the stopping criterion which depends on the CEI changes between iterations is only fulfilled at iteration 5.
Therefore, the algorithm needs five iterations to fulfill the stopping criterion, 26 experiments for finding the critical gains and the initial set of 15 experiments for training of the GP model of the cost, resulting in 46 experiments in total. The resulting performance is significantly improved, as shown in Figure \ref{fig:cost}. All performance metrics are improved, especially the initial overshoot $C_{ST}$ shows a decrease of 50 \%. The frequency domain plot shows that the found parameters are stable without introducing vibrations (e.g. high-frequency components remain low), and exemplifies the decrease in the low-frequency range, corresponding to improved tracking.

Upon re-tuning only a small number of experiments will be needed to account for changes in the system. The old experiments can still be used as prior information.

\begin{figure}[ht]
\centering
\includegraphics[width=0.45\textwidth]{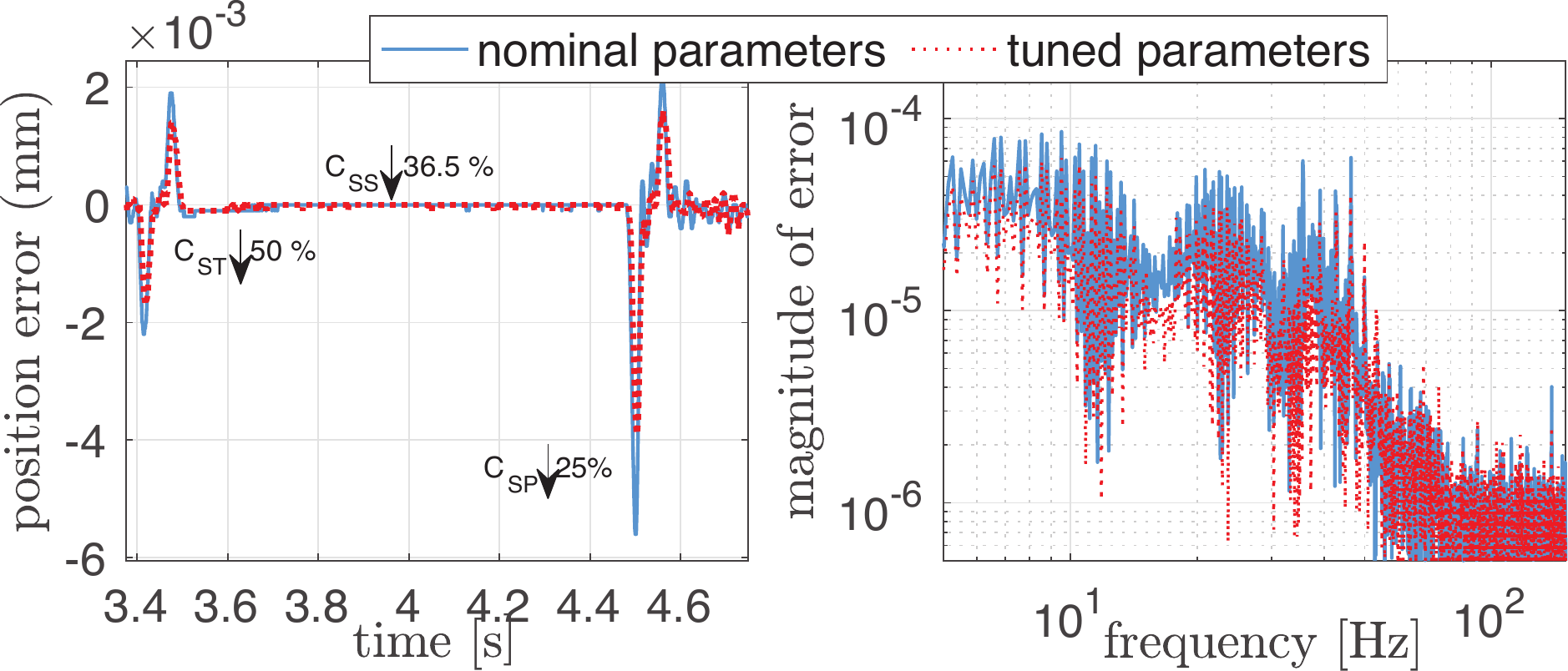}
\caption{The position error signal and the improvement of each metric before and after tuning, in time-domain (left), and the position error signal in frequency domain (right).}
\label{fig:cost}
\end{figure}

\section{Conclusion}

Constrained Bayesian optimization has been applied to tune a cascaded controller in a model-free approach, while ensuring safety  by  experimentally determining  the  critical gains  of the controller. 
Scanning in advance to determine the critical gains provides the means to inform the Bayesian optimization about safety and stability bounds in the system. The proposed algorithm can run autonomously, between production cycles, without disturbing the operation of the system.
If the system undergoes significant physical changes between tuning cycles, the pre-learned critical values might not be accurate. The proposed algorithm can then be extended to accommodate such changes by incorporating task (context) parameters in the modeled system performance.  This extension enables the transfer of the prior models instead of re-learning them from scratch.  
With the current method, the achieved performance exceeded the nominal one by more than 20\%, while automating and standardizing the tuning procedure.
\vspace{-2mm}

\iftrue
\section*{Appendix: Optimization of the Acquisition Function}
\label{sec_simulation_2}
The main part of the proposed approach is the sampling step performed by finding the optimum of the acquisition function in \eqref{eqn:x_next}. 
This can be done by various methods including nonlinear programming techniques such as limited-memory BFGS \cite{Frazier}, \cite{Liu},  grid search \cite{Gardner}, and global optimization methods like particle swarm optimization \cite{Khosravi}. 
The main drawback of nonlinear programming is being prone to entrapment in local minima when the optimization problem is non-convex, like in  \eqref{eqn:x_next}. 
While the grid search methods have the potential of finding the global optima, the grid evaluation depends strongly on the resolution and the dimension of the grid. 
Finally, the particle swarm optimization (PSO) is a global optimization method that can be adapted to constraints, works in continuous space, and scales well with the dimensionality of the problem \cite{Kennedy}.
Table \ref{tab:PSO_comp} shows a comparison of the PSO method using 10 particles with a grid-based evaluation for different resolutions of the grid in the numerical CBO experiments following the cost without safety penalization for 10 repetitions. 
Each trial starts with the same initial set of 25 points, drawn by Latin hypercube sampling. 
For all experiments the parameters $\Kp$, $\Kv$ and $\Ti$ are optimized simultaneously. 
For grid maximization, the values are discretized according to the resolution of the grid. 
When a fine resolution is used, the PSO algorithm is computationally faster and terminates closer to the optimum with a higher number of iterations. 
Note that the number of iterations is balanced with the final cost value through the termination criterion \eqref{eq:stop_aCEI}. 
Solving  \eqref{eqn:x_next} using a coarse grid optimization achieves smaller number of iterations and less computational time comparing to PSO, again with higher cost compared to the PSO result. 
It is empirically observed that improving the resolution of the grid by factor two in all three dimensions increases the computational time by factor 6, while the PSO is not limited to a grid resolution.
On other hand, regarding the performance of PSO with larger swarm population, similar results are obtained when 100 particles are used.
\begin{table} [h]
\caption{Comparison of PSO and grid search for acquisition function optimization (mean and 95\% confidence interval)}
\label{tab:PSO_comp}
    \centering
    \ra{1.3}
    \begin{tabular}{ @{}c c c c c @{} }\toprule
        & median & median & \\
     search method & num. of iter. $N$ & comp. time~$t$ [s] & cost $f$\\
     \midrule
     grid search & & &\\
     240$\times$75$\times$4 & $39$ & $45.2$ & $1.47\text{e-}4\pm1.01\text{e-}5$\\\midrule
     grid search & & &\\
     480$\times$150$\times$8 & $32$ & $172.5$ & $1.46\text{e-}4\pm 1.1\text{e-}5$\\\midrule
     PSO 10 particles & $49.5$ & $115.84$ & $1.45\text{e-}7\pm 9.63\text{e-}6$\\
   \bottomrule
    \end{tabular}
\end{table}
\begin{figure}[t]
\centering
  \includegraphics[width=0.4\textwidth]{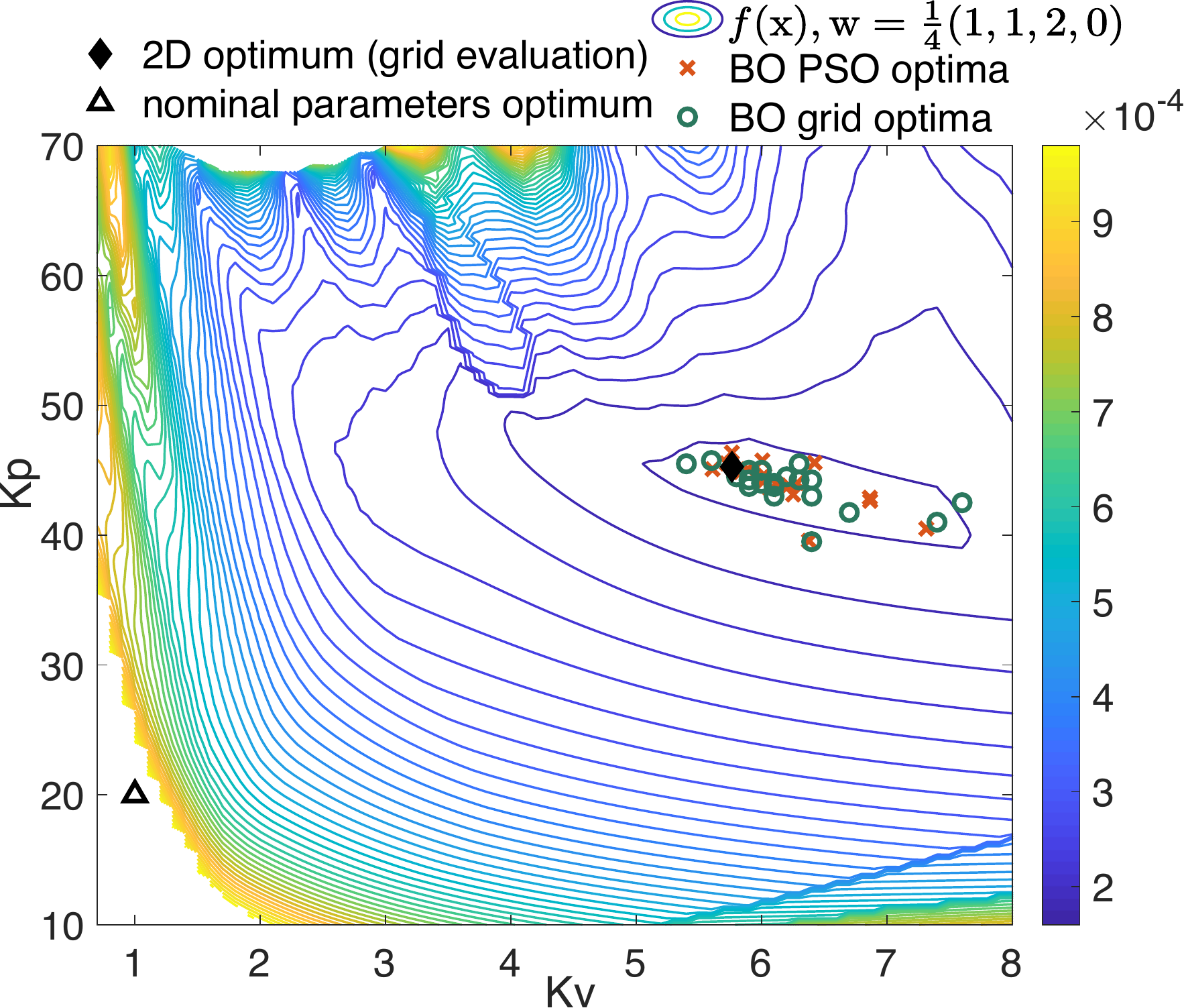}
\caption{Comparison of grid search and PSO to maximize the acquisition function}
\label{fig:GX_sim_grid_pso_comp}
\end{figure}
Figure \ref{fig:GX_sim_grid_pso_comp} shows that the optimized gains are spread in the vicinity of the the optimal controller parameters.
This typical feature of the proposed tuning is due to the flat shape of the graph of the cost function over this region.
The stopping criterion can be modified to tighten the final allowed optima while balancing the number of the required iterations for reaching to the optimal gains.

\fi
\bibliography{bibliography.bib}

\begin{thebibliography}{10}
\providecommand{\url}[1]{#1}
\csname url@samestyle\endcsname
\providecommand{\newblock}{\relax}
\providecommand{\bibinfo}[2]{#2}
\providecommand{\BIBentrySTDinterwordspacing}{\spaceskip=0pt\relax}
\providecommand{\BIBentryALTinterwordstretchfactor}{4}
\providecommand{\BIBentryALTinterwordspacing}{\spaceskip=\fontdimen2\font plus
\BIBentryALTinterwordstretchfactor\fontdimen3\font minus
  \fontdimen4\font\relax}
\providecommand{\BIBforeignlanguage}[2]{{%
\expandafter\ifx\csname l@#1\endcsname\relax
\typeout{** WARNING: IEEEtran.bst: No hyphenation pattern has been}%
\typeout{** loaded for the language `#1'. Using the pattern for}%
\typeout{** the default language instead.}%
\else
\language=\csname l@#1\endcsname
\fi
#2}}
\providecommand{\BIBdecl}{\relax}
\BIBdecl

\bibitem{borase2020review}
R.~P. Borase, D.~Maghade, S.~Sondkar, and S.~Pawar, ``A review of {PID}
  control, tuning methods and applications,'' \emph{International Journal of
  Dynamics and Control}, pp. 1--10, 2020.

\bibitem{Astrom}
K.~Åström and T.~Hägglund, ``Automatic tuning of simple regulators with
  specifications on phase and amplitude margins,'' \emph{Automatica}, vol.~20,
  pp. 645--651, 09 1984.

\bibitem{Hang3}
C.~C. {Hang}, A.~P. {Loh}, and V.~U. {Vasnani}, ``Relay feedback auto-tuning of
  cascade controllers,'' \emph{IEEE Transactions on Control Systems
  Technology}, vol.~2, no.~1, pp. 42--45, March 1994.

\bibitem{Kennedy}
J.~Kennedy and R.~Eberhart, ``Particle swarm optimization,''
  \emph{International Conference on Neural Networks}, pp. 1942--1948, 1995.

\bibitem{Dorigo}
M.~Dorigo, M.~Birattari, and T.~Stutzle, ``Ant colony optimization,''
  \emph{IEEE Computational Intelligence Magazine}, vol.~1, no.~4, pp. 28--39,
  2006.

\bibitem{Mitchell}
M.~Mitchell, \emph{An Introduction to Genetic Algorithms}.\hskip 1em plus 0.5em
  minus 0.4em\relax MIT Press, 1998.

\bibitem{Solihin}
M.~Solihin, L.~Tack, and L.~K. Moey, ``Tuning of {PID} controller using
  particle swarm optimization,'' \emph{International Conference on Advanced
  Science, Engineering and Information Technology}, vol.~1, 01 2011.

\bibitem{Chiha}
I.~Chiha, N.~Liouane, and P.~Borne, ``Tuning {PID} controller using
  multiobjective ant colony optimization,'' \emph{Applied Computational
  Intelligence and Soft Computing}, vol. 2012, 2012.

\bibitem{Zhang2006}
{Ping Zhang}, {Mingzhe Yuan}, and {Hong Wang}, ``Self-tuning {PID} based on
  adaptive genetic algorithms with the application of activated sludge aeration
  process,'' in \emph{2006 6th World Congress on Intelligent Control and
  Automation}, vol.~2, 2006, pp. 9327--9330.

\bibitem{Li2019}
L.~Li, Y.~Liu, L.~Li, and J.~Tan, ``Kalman-filtering-based iterative
  feedforward tuning in presence of stochastic noise: With application to a
  wafer stage,'' \emph{IEEE Transactions on Industrial Informatics}, vol.~15,
  no.~11, pp. 5816--5826, 2019.

\bibitem{IterTuning_Li2018}
X.~Li, S.-L. Chen, J.~Ma, C.~S. Teo, and K.~K. Tan, ``Data-driven model-free
  iterative tuning approach for smooth and accurate tracking,'' in \emph{2018
  IEEE/ASME International Conference on Advanced Intelligent Mechatronics
  (AIM)}, 2018, pp. 593--598.

\bibitem{VarGain_Li2015}
M.~Li, Y.~Zhu, K.~Yang, and C.~Hu, ``A data-driven variable-gain control
  strategy for an ultra-precision wafer stage with accelerated iterative
  parameter tuning,'' \emph{IEEE Transactions on Industrial Informatics},
  vol.~11, no.~5, pp. 1179--1189, 2015.

\bibitem{FeedforLearning_Song2020}
F.~Song, Y.~Liu, W.~Jin, J.~Tan, and W.~He, ``Data-driven feedforward learning
  with force ripple compensation for wafer stages: A variable-gain robust
  approach,'' \emph{IEEE Transactions on Neural Networks and Learning Systems},
  pp. 1--15, 2020.

\bibitem{Constrained_Radac2013}
M.-B. Radac, R.-E. Precup, S.~Preitl, C.-A. Dragos, and E.~M. Petriu,
  ``Constrained data-driven controller tuning for nonlinear systems,'' in
  \emph{IECON 2013 - 39th Annual Conference of the IEEE Industrial Electronics
  Society}, 2013, pp. 3404--3409.

\bibitem{Khosravi2020}
M.~{Khosravi}, V.~{Behrunani}, R.~S. {Smith}, A.~{Rupenyan}, and J.~{Lygeros},
  ``{Cascade Control: Data-Driven Tuning Approach Based on {B}ayesian
  Optimization},'' \emph{IFAC world congress 2020}, p. arXiv:2005.03970, 2020.

\bibitem{Khosravi2020b}
M.~Khosravi, V.~Behrunani, P.~Myszkorowski, R.~S. Smith, A.~Rupenyan, and
  J.~Lygeros, ``Performance-driven cascade controller tuning with {B}ayesian
  optimization,'' \emph{IEEE Transactions on Industrial Electronics}, pp. 1--1,
  2021.

\bibitem{Gardner}
J.~Gardner, M.~Kusner, Zhixiang, K.~Weinberger, and J.~Cunningham, ``{B}ayesian
  optimization with inequality constraints,'' in \emph{International Conference
  on Machine Learning}, vol.~32, 2014, pp. 937--945.

\bibitem{Maier}
M.~Maier, R.~Zwicker, M.~Akbari, A.~Rupenyan, and K.~Wegener, ``{B}ayesian
  optimization for autonomous process set-up in turning,'' \emph{CIRP Journal
  of Manufacturing Science and Technology}, 2019.

\bibitem{MaierGrind2020}
M.~Maier, A.~Rupenyan, C.~Bobst, and K.~Wegener, ``Self-optimizing grinding
  machines using {G}aussian process models and constrained {B}ayesian
  optimization,'' \emph{The International Journal of Advanced Manufacturing
  Technology}, vol. 108, pp. 528--552, May 2020.

\bibitem{BOcontrolRun2020}
E.~{Moya-Lasheras} and C.~{Sagues}, ``Run-to-run control with {B}ayesian
  optimization for soft landing of short-stroke reluctance actuators,''
  \emph{IEEE/ASME Transactions on Mechatronics}, pp. 1--1, 2020.

\bibitem{Sui}
Y.~Sui, A.~Gotovos, J.~W. Burdick, and A.~Krause, ``Safe exploration for
  optimization with {G}aussian processes,'' in \emph{International Conference
  on Machine Learning (ICML)}, 2015.

\bibitem{Berkenkamp}
F.~Berkenkamp, A.~Krause, and A.~P. Schoellig, ``{B}ayesian optimization with
  safety constraints: Safe and automatic parameter tuning in robotics,''
  \emph{CoRR}, vol. abs/1602.04450, 2016.

\bibitem{Berkenkamp2}
F.~Berkenkamp, A.~P. Schoellig, and A.~Krause, ``Safe controller optimization
  for quadrotors with {G}aussian processes,'' \emph{arXiv:1509.01066}.

\bibitem{Khosravi}
M.~{Khosravi}, A.~{Eichler}, N.~{Schmid}, R.~S. {Smith}, and P.~{Heer},
  ``Controller tuning by {B}ayesian optimization an application to a heat
  pump,'' \emph{European Control Conference}, pp. 1467--1472, June 2019.

\bibitem{Duivenvoorden}
R.~R. Duivenvoorden, F.~Berkenkamp, N.~Carion, A.~Krause, and A.~P. Schoellig,
  ``Constrained {B}ayesian optimization with particle swarms for safe adaptive
  controller tuning,'' \emph{IFAC-PapersOnLine}, vol.~50, no.~1, pp. 11\,800 --
  11\,807, 2017, 20th IFAC World Congress.

\bibitem{Frazier}
P.~I. Frazier, ``A tutorial on {B}ayesian optimization,'' 2018.

\bibitem{Nguyen}
V.~Nguyen, S.~Gupta, S.~Rana, C.~Li, and S.~Venkatesh, ``Regret for expected
  improvement over the best-observed value and stopping condition,''
  \emph{Proceedings of the Ninth Asian Conference on Machine Learning},
  vol.~77, pp. 279--294, 15-17 Nov 2017.

\bibitem{mckay2000comparison}
M.~D. McKay, R.~J. Beckman, and W.~J. Conover, ``A comparison of three methods
  for selecting values of input variables in the analysis of output from a
  computer code,'' \emph{Technometrics}, vol.~42, pp. 55--61, 2000.

\bibitem{Rasmussen}
C.~E. Rasmussen, ``{G}aussian processes for machine learning.''\hskip 1em plus
  0.5em minus 0.4em\relax MIT Press, 2006.

\bibitem{Rohrig}
C.~Rohrig and A.~Jochheim, ``Identification and compensation of force ripple in
  linear permanent magnet motors,'' \emph{Proceedings of the 2001 American
  Control Conference}, vol.~3, pp. 2161--2166, June 2001.

\bibitem{Villegas}
F.~Villegas, R.~Hecker, M.~Peña, D.~Vicente, and G.~Flores, ``Modeling of a
  linear motor feed drive including pre-rolling friction and aperiodic cogging
  and ripple,'' \emph{The International Journal of Advanced Manufacturing
  Technology}, vol.~73, 07 2014.

\bibitem{Hang2}
C.~Hang, K.~{\AA}str{\"o}m, and W.~Ho, ``Relay auto-tuning in the presence of
  static load disturbance,'' \emph{Automatica}, vol.~29, pp. 563 -- 564, 1993.

\bibitem{Liu}
D.~C. Liu and J.~Nocedal, ``On the limited memory {BFGS} method for large scale
  optimization,'' \emph{Mathematical Programming}, vol.~45, no.~1, pp.
  503--528, Aug 1989.

\end{thebibliography}
\bibliographystyle{IEEEtran}
\end{document}